\documentclass[twocolumn]{aastex631}

\begin{document}

\title{The central engine of low-luminosity radio galaxy 3C~270 (NGC~4261)}

\author[0009-0008-2985-0726]{Chengzhang Jiang}
\affiliation{Department of Physics and Astronomy, University of California, Riverside, 900 University Ave, Riverside CA 92521, USA}

\author{Robert Antonucci}
\affiliation{Department of Physics, University of California, Santa Barbara, CA 93106-9530, USA}

\author[0000-0002-0433-9656]{J. A. Acosta-Pulido}
\affiliation{Instituto de Astrof\'isica de Canarias, Calle V\'ia L\'actea, s/n, E-38205, La Laguna, Tenerife, Spain}
\affiliation{Departamento de Astrof\'isica, Universidad de La Laguna, E-38206, La Laguna, Tenerife, Spain}

\author[0000-0002-3471-981X]{Patrick Ogle}
\affiliation{Space Telescope Science Institute Baltimore, MD 21218, USA}

\correspondingauthor{Chengzhang Jiang}
\email{cjian079@ucr.edu}
 
\begin{abstract}
We present the polarization spectra of the nucleus of 3C~270. We confirm that the polarization angle of both the continuum and the emission lines are close to perpendicular to the jet direction after careful correction of interstellar polarization, which indicates polar scattering. The Stokes flux spectrum resembles the total flux spectrum, with no need for a broad component from the Broad Line Region. Over 94\% of a sample of Seyfert I broad line profile would be significantly detected if present in our polarized flux spectrum. We favor the hypothesis that we are observing the continuum including any contribution from Radiatively Inefficient Accretion Flow and beamed synchrotron instead of a Big Blue Bump, as well as the innermost Narrow Line Region, through reflection. This makes 3C~270 the third known case, after NGC~4258 and Centaurus~A, where only narrow lines (and the continuum, if present) are scattered, with no evidence of an underlying Big Blue Bump.

\end{abstract}


\section{Introduction} 
\label{sec:intro}
The nearby elliptical galaxy NGC 4261 ($z = 0.0074$) hosts a low-luminosity Active Galactic Nuclei (AGN) and also a bright radio source known as 3C~270. This galaxy is renowned for its prominent geometrically thick dust disc \citep{Jaffe1993} rotating around a supermassive black hole with a mass $M_\mathrm{BH}$ of $1.62\,\pm\,0.04\,\times\,10^{9}\,M_{\odot}$ , as reported by \citet{Boizelle2021} and \cite{Ruffa2023}. 

This famous FR~I type radio galaxy has been the subject of numerous studies trying to establish its accretion mode : radiatively efficient, "thermal" vs. radiatively inefficient, "non-thermal" \citep{Antonucci2012}. This interest is motivated by the fact that 3C~270 appears to be a low-luminosity active galaxy showing puzzling evidence for significant accretion power, making it an ideal testing ground for our current understanding of accretion onto supermassive black holes at low Eddington ratios. On one hand, the low-efficiency accretion is supported by emission-line diagnostics from ground-based spectra, which classify 3C~270 as a low-ionization nuclear emission-line region (LINER) \citep{Ho1995, Jaffe1996}. However, \textit{Hubble Space Telescope} (\textit{HST}) spectra show Seyfert-type narrow-line ratios on nuclear scales \citep{Ferrarese1996}, indicating the presence of a hard radiation field typical for radiatively efficient accretion \footnote{The nucleus predominantly exhibits a LINER spectrum, but space-based observations within a central 0.1$^{\prime\prime}$ aperture reveal nuclear line ratios ([O~III]\,/\,H$\beta$\,$\simeq$\,5) that fall in the Seyfert region on the [S~II] and [N~II] BPT diagrams \citep{Kewley2006, Kauffmann2003}, suggesting the possible presence of a hard radiation field. Old stellar population \citep{Wills2004} with a low star formation rate \citep[$\lesssim$\,0.01\,M$_\odot$\,yr$^{-1}$,][]{Vaddi2016} indicates that underlying Balmer absorption from host galaxy \citep[$\lesssim$\,1.5\AA,][]{Worthey1994} does not significantly affect the line ratios (Equivalent Width of H$\beta\gtrsim$\,10\AA) on the BPT diagram.}.

From \textit{Spitzer} spectroscopy, \citet{Leipski2009} showed that 3C~270 is dominated by warm thermal dust ($\sim$\,240\,K) in the mid-infrared, too warm for stellar-heated dust, and indicative of a possible hidden thermal AGN \citep{Ogle2006}. Very Large Telescope  Imager and Spectrometer for mid Infrared (VLT-VISIR) data at 12 micron reveal a matching point source, which proves that the warm thermal dust has indeed a nuclear origin \citep[$<$\,60\,pc;][]{Asmus2011}. However, in some clumpy torus models, a powerful nuclear thermal dust component is not expected for sources of low luminosity. Instead, the models predict that the torus (and the broad line region) vanishes at low bolometric luminosity ($\sim$\,10$^{42}$\,erg\,s$^{-1}$) with 3C~270 having $\sim$\,$1.7 \times 10^{42}$\,erg\,s$^{-1}$ \citep{Ho1999} and/or at low Eddington ratios \citep{Honig2007, Elitzur2008, Elitzur2009, Gonzalez-Martin2017}. Additionally, for the very low Eddington ratio of 3C~270 \citep[$L / L_\mathrm{edd} \sim 10^{-5}$, ][]{Ho1999}, theory predicts advection-dominated accretion \citep{Esin1997}, thus limiting the supply of hard ionizing radiation. However, the nuclear optical emission lines in 3C~270 requires a central engine to produce sufficient ionizing photons. Furthermore, the reprocessed light observed as nuclear warm dust emission necessitates a radiation field sufficient to power the emission. These discrepancies illustrate the need to establish more accurate observational constraints to the models.

One important diagnostic that could place strong constraints on the models for low-luminosity activity is the detection of broad emission lines and the Big Blue Bump continuum component. These are generally considered reliable probes for the presence of a radiatively efficient accretion flow, and the detection of broad lines in polarized light would be the most convincing proof for a hidden thermal AGN. Spectropolarimetry has been used successfully many times to reveal hidden broad lines in narrow-line objects \citep[e.g.,][]{Antonucci1984, Ogle1997, Ramos-Almeida2016}. Additionally, 3C~270 has been observed spectropolarimetrically by \citet{Barth1999}, surprisingly revealing possibly broadened H$\alpha$ in polarized light. Unfortunately, the low signal-to-noise ratio (SNR) precluded a secure detection and width measurement of the apparent emission line feature. In concert with the detection of warm nuclear dust emission, establishing the presence of broad emission lines in 3C~270 would provide strong constraints for our current models of low-luminosity activity, which predict both of these features to be absent. It should be noted however that the theoretical expectations are already known to be contradicted in some cases (e.g., NGC~1052, \citealp{NGC1052}).

In this work, we present a deeper optical spectropolarimetric observation on 3C~270, aiming to disentangle the H$\alpha$ complex in search of hidden broad line. In Section \ref{sec:observation}, we describe the observations and data reduction and careful removal of interstellar polarization. In Section \ref{sec:results}, we analyze different models for the polarized flux spectrum. In Section \ref{sec:discussion}, we discuss the implication of accretion mode as from the polarized flux spectrum and make a summary. Throughout the paper, we adopt a $\Lambda$CDM cosmology with $\Omega_{\Lambda}$ = 0.7, $\Omega_{m}$ = 0.3, and H$_{0}$ = 70\,km\,s$^{-1}$\,Mpc$^{-1}$.

\section{Observations and Data Reduction}
\label{sec:observation}
\begin{figure*}
    \centering
    \includegraphics[width=0.9\linewidth]{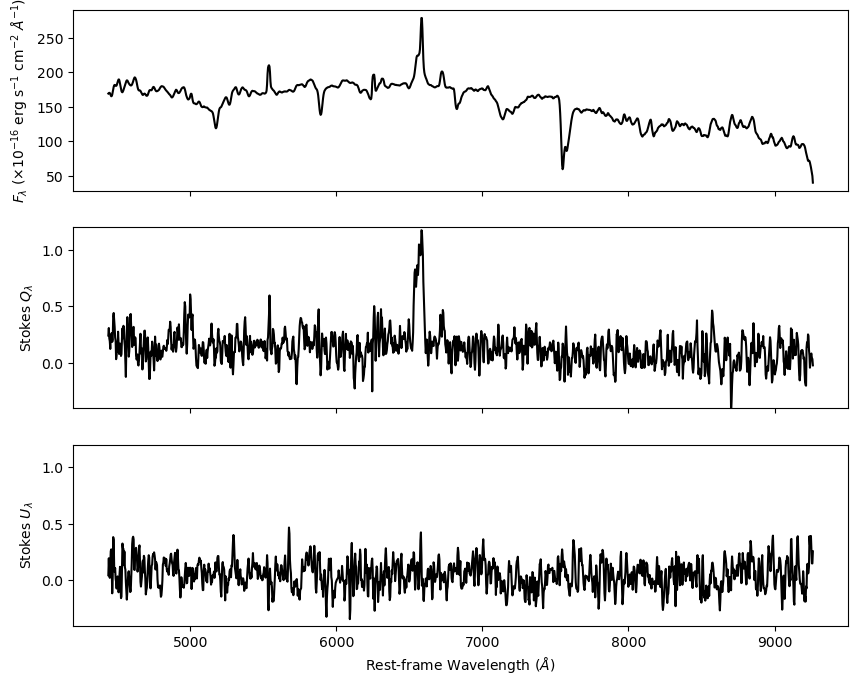}
    \caption{The figure displays the total flux spectrum along with the unnormalized Stokes $Q$ and $U$ spectra before interstellar polarization correction. All three spectra have the same units ($\times 10^{-16}$\,erg\,cm$^{-2}$\,s$^{-1}$\,\AA$^{-1}$).}
    \label{fig:IQU}
\end{figure*}
The observations of 3C~270 were conducted using the FOcal Reducer and low dispersion Spectrograph (FORS2), mounted on the Unit Telescope 1 of the Very Large Telescope (VLT) at the European Southern Observatory (ESO). The observations were spread over three nights: 2014\,/\,01\,/\,24, 2014\,/\,02\,/\,28, and 2014\,/\,03\,/01 (Program ID: 092.B-0401A). The seeing conditions vary within the three nights as follows: 0.9$^{\prime\prime}$~-~1.1$^{\prime\prime}$ (2014\,/\,01\,/\,24); 1.2$^{\prime\prime}$~-~1.7$^{\prime\prime}$ (2014\,/\,02\,/\,28); 0.5$^{\prime\prime}$~-~0.9 $^{\prime\prime}$(2014\,/\,03\,/\,01). All data were taken in PMOS mode with the 300V grism.

The observations employed the Half-Wave Plate (HWP) at four different position angles of 0.0$^{\circ}$, 22.5$^{\circ}$, 45.0$^{\circ}$, and 67.5$^{\circ}$ \footnote{The HWP angles correspond to 0.0$^{\circ}$, 45.0$^{\circ}$, 90.0$^{\circ}$, and 135.0$^{\circ}$ on the sky plane.}. A Wollaston prism was used to split the light into ordinary ray (o-ray) and extraordinary ray (e-ray). Due to the dual-beam design of FORS2, both o-ray and e-ray were recorded simultaneously, with a separation of about 22 arcseconds. The total on-source exposure time for each half-wave plate position was 3000s.

The data were reduced using routines written in IDL, following the procedures described by \citet{Martinez2015} and \citet{Ramos-Almeida2016}. The steps included bias subtraction, flat-field correction, wavelength calibration, sky subtraction, cosmic ray removal, and flux calibration. Both o-ray and e-ray spectra are extracted using an aperture with a radius equal to the FWHM of the seeing for each epoch, and then combined with weights based on the exposure time of each epoch. The o-ray and e-ray spectra were then combined to compute the normalized Stokes parameters $q$ and $u$ using the following formulas:
\begin{equation}
q = \frac{1}{2} \left( \frac{f_{o,0} - f_{e,0}}{f_{o,0} + f_{e,0}} - \frac{f_{o,45} - f_{e,45}}{f_{o,45} + f_{e,45}} \right)
\end{equation}
\begin{equation}
u = \frac{1}{2} \left( \frac{f_{o,22.5} - f_{e,22.5}}{f_{o,22.5} + f_{e,22.5}} - \frac{f_{o,67.5} - f_{e,67.5}}{f_{o,67.5} + f_{e,67.5}} \right)
\end{equation}
where $f$ is the measured intensity, with the first subscript indicating the ordinary (o) or extraordinary (e) ray, and the second subscript specifying the HWP rotation angle.  

The total intensity $I$ is computed as: $I = (f_{o,0} + f_{e,0} + f_{o,45} + f_{e,45} + f_{o,22.5} + f_{e,22.5} + f_{o,67.5} + f_{e,67.5}) / 4$.

The linear polarization degree $P$ and polarization angle PA are derived from $q$ and $u$ using the standard relations:  
\begin{equation}
P = \sqrt{q^2 + u^2}
\end{equation}
\begin{equation}
\mathrm{PA} = \frac{1}{2} \arctan \left( \frac{u}{q} \right).
\end{equation}

For calibration, we selected a zero-polarization standard star (HD~109055) and a polarized standard star (Vela1~95), both observed on 2024\,/\,01\,/\,24 after the first source observation. Flux and polarimetric standards were part of the default nightly calibration plan. 

In the R band, Vela1~95 has a polarization degree of $7.89 \pm 0.04\%$ and a polarization angle of $172.1^\circ \pm 0.2^\circ$\footnote{\url{https://www.eso.org/sci/facilities/paranal/instruments/fors/inst/pola.html}}. For comparison with the standard, we integrate the normalized Stokes parameters $q$ and $u$ from our data, weighted by the Johnson R filter transmission. From our measurements, the polarization degree is 7.79\,$\pm$\,0.04\% with a polarization angle of $172.9^\circ$\,$\pm$\,$0.2^\circ$, very close to the standard values. Using the same procedure, we measure the polarization degree of HD~109055, finding 0.06\,$\pm$\,0.04\%, consistent with expected null polarization. The close agreement between our measured and tabulated values indicates that the instrumental polarization in the R band is $\lesssim$\,$0.06\%$.

The total flux spectrum as well as the Stokes $Q$ and $U$ spectra before interstellar polarization correction are shown in Fig.~\ref{fig:IQU}.

\subsection{Interstellar Polarization}
Since the continuum polarization is quite low, interstellar polarization (ISP) may significantly impact our interpretation of its origin. \citet{Serkowski1975} demonstrate that the wavelength-dependent polarization degree of the ISP can be accurately modeled using two parameters, $P_\mathrm{max}$ and $\lambda_\mathrm{max}$. With the additional constraint of the ISP polarization angle, we can derive the complete ISP information, including the normalized (fractional) Stokes parameters $q$ and $u$. Key parameters for the ISP can be estimated from polarization measurements of stars close to the central source, which provides a local basis for ISP correction.

We start with the polarization catalog of Galactic stars by \citet{Heiles2000} and take the polarization properties of the closest star to the source to serve as a proxy for the ISP of the source. We also require the proxy star to be sufficiently distant from the Earth to maximally sample the ISP caused by Galactic dust along the line of sight. The best candidate matching these criteria is HD~107289, located $\sim$ \,300\,pc away from the Earth at Galactic latitude of $\sim 67^{\circ}$ and 12$^{\prime}$ away from 3C~270 in projection. Galactic ISP exhibits intrinsic asymmetry on large scale \citep{Berdyugin2014}. However, their Fig.~3 shows that the polarization angle of ISP is largely coherent around 3C~270. We verify this coherence by comparing the polarization angle of HD~107289 to that of the nearby stars within 5$^\circ$, finding variations within ~10$^\circ$, which will not affect our analysis. Therefore, HD~107289 is a good proxy for measuring Galactic ISP. The measured polarization degree is 0.084\%\,$\pm$\,0.012\% and the polarization angle is $64^{\circ}$\,$\pm$\,$4.1^{\circ}$. To check for residual dust between HD~107289 and 3C~270 , we use the 3D dust map from \citet{Green2019} to calculate the integrated extinction along the line of sight. In the direction of 3C~270, the integrated Galactic extinction is $A_V$\,=\,0.059\,$\pm$\,0.014\,mag, while toward HD~107289, the value is 0.047\,$\pm$\,0.018\,mag. This indicates that there is minimal dust beyond HD~107289 that would contribute to ISP within our Galaxy. Taking $P_\mathrm{max}$\,=\,0.084\%, $\mathrm{PA}$\,=\,64$^{\circ}$ and assuming $\lambda_\mathrm{max}$ is 5500\,\AA, we can then construct the synthetic normalized $q$ and $u$ spectra as follows:

\begin{equation}
    q_\mathrm{ISP}(\lambda) = P(\lambda) \times \cos(2\mathrm{PA}) \\
\end{equation}
\begin{equation}
    u_\mathrm{ISP}(\lambda) = P(\lambda) \times \sin(2\mathrm{PA})
\end{equation}

\begin{figure}
    \centering
    \includegraphics[width=0.9\linewidth]{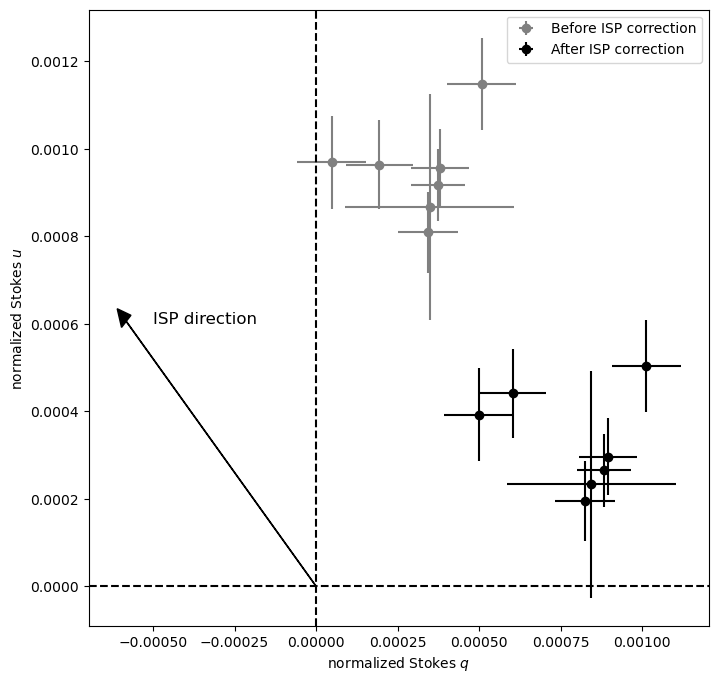}
    \caption{The black points indicate the normalized Stokes $q$ and $u$ of the source continuum at different wavelength bins after ISP correction, while the gray points show the values before correction. The arrow denotes the direction of the interstellar polarization vector, the length of which is set to $P_\mathrm{max}$.}
    \label{fig:normalized_qu}
\end{figure}

The computed normalized Stokes parameters are then subtracted from the observed normalized $q$ and $u$ to derive the intrinsic values. The error of the intrinsic normalized Stokes parameters is calculated by combining the ISP error and the measurement error in quadrature. The normalized Stokes parameter $q$ and $u$ are shown in Fig.~\ref{fig:normalized_qu}.

\section{Results}
\label{sec:results}
\begin{figure*}
    \centering
    \includegraphics[width=0.9\linewidth]{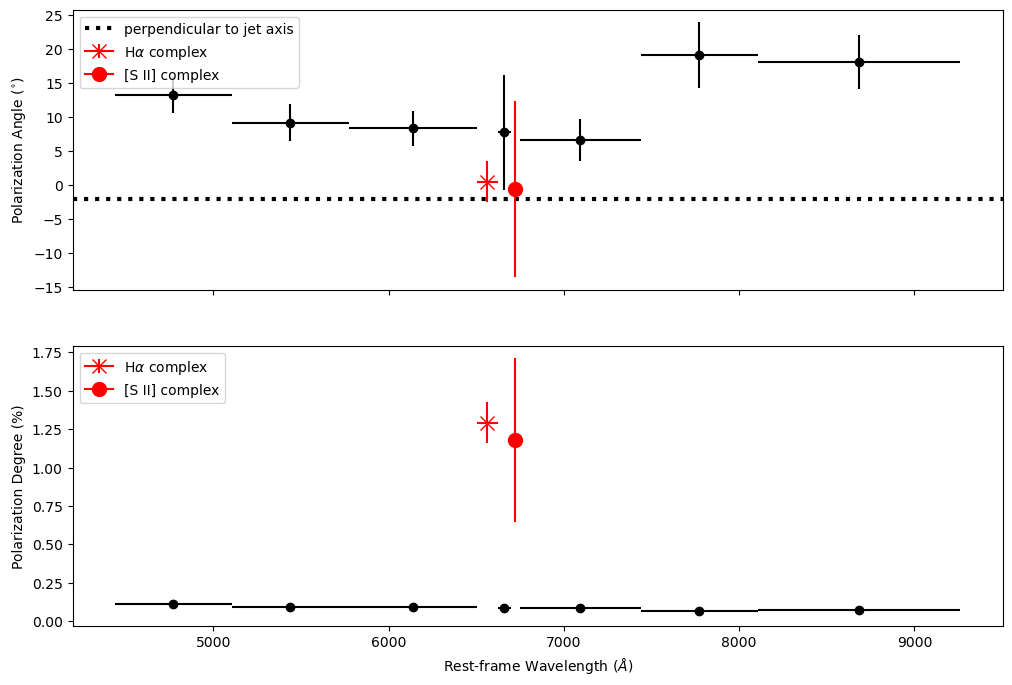}
    \caption{The upper panel presents the polarization angle as a function of wavelength. Black points indicate the continuum polarization. The red cross represents the integrated polarization of the H$\alpha$ complex after continuum subtraction and the red dot denotes the integrated polarization of the [S II] complex, also following continuum subtraction. The black dotted line marks the position angle perpendicular to the jet axis. The lower panel follows the same configuration as the upper panel but displays the polarization degree as a function of wavelength.}
    \label{fig:pa_pd}
\end{figure*}

The radio jet position angle is measured $88^\circ$\,$\pm$\,$1^{\circ}$ at 4.885\,GHz \citep{Birkinshaw1985}. After correcting for the ISP, we find that the polarization angle of the continuum is close to perpendicular to the jet direction, as shown in Fig.~\ref{fig:pa_pd} (see also Table \ref{tab:line_polarization}). At wavelength of $\sim $\,6000\,\AA, we found a continuum polarization angle of $8.4^{\circ}$\,$\pm$\,$2.6^{\circ}$, aligning well with the findings of \citet{Capetti2007} at matching wavelength. Their broadband polarimetry, primarily tracing continuum polarization, shows a polarization angle of $-7.5^{\circ}$\,$\pm$\,$5.6^{\circ}$, perpendicular to the jet. Most importantly, the polarization angle of both the H$\alpha$ and [S~II] complexes, which are less subject to ISP correction, are $0.5^{\circ}$\,$\pm$\,$3.0^{\circ}$ and $-0.6^{\circ}$\,$\pm$\,$12.9^{\circ}$, very closely perpendicular to the jet axis at 88$^{\circ}$. Our results are in excellent agreement with theirs, demonstrating the precision of our ISP correction as well.

\citet{Capetti2007} report the optical polarization of a sample of FR~I radio galaxies. In particular, their Table 2 shows an overall uncorrelated polarization angles to the jet position angle. The lack of consistent perpendicularity across their sample leads \citet{Capetti2007} to attribute the observed polarization to intrinsic radiation mechanisms, most likely synchrotron emission. However, a notable distinction between 3C~270 and the other FR~Is in \citet{Capetti2007}, besides its perpendicular polarization, is the high X-ray column density of $N_\mathrm{H} \gtrsim 7 \times 10^{22}$\,cm$^{-2}$ \citep{Zezas2005, Worrall2010}. Thus, 3C~270 is similar to the FR~I prototype Centaurus~A, which wasn't observed by \citet{Capetti2007}, but which also has a perpendicular polarization and a high column density associated with the nuclear region, two signatures of a hidden AGN (Marin et al. 2024). It's also worth noting that the polarization angle is constant in these two objects over time. Again this argues against the notion that these two FR~Is show their synchrotron cores directly in the optical. The case is clinched in Centaurus~A by the similar high perpendicular polarization in the line complex. In 3C~270, we also see that both H$\alpha$ and [S~II] complex are polarized at the same polarization angle, though direct line emission must dilute the scattered emission substantially in the latter case since the polarization is only $\sim$\,0.1\%.

In Fig.~2 of \citet{Capetti2007}, the polarization degree is plotted as a function of aperture radius to investigate the concentration of polarization. The polarization degree of 3C~270 reaches as high as $\sim$\,10\% at the central pixel, consistent with a point source origin, and drops to $\lesssim$\,0.2\% at an aperture radius of $\sim$\,$1.2^{\prime\prime}$. Given that our observation uses a comparable aperture size to the latter, this polarization degree is broadly consistent with our measured continuum polarization.

A key objective of our observations is to determine whether broad H$\alpha$ emission, coupled with the Big Blue Bump continuum, can be detected in scattered (polarized) light, as predicted by the unification model. We confirm the detection of polarized narrow emission lines (the H$\alpha$ complex) with a polarization angle perpendicular to the jet axis by \citet{Barth1999}. The alignment of both the emission lines and the continuum polarization perpendicular to the jet in our data strengthen a scattering scenario, similar to what has been observed in Centaurus~A. We now examine the compositions of the emission line complex in the polarized flux spectrum in detail.

\begin{table}
    \caption{Wavelength dependence of polarization}
    \label{tab:line_polarization}      
    \centering                                      
    \begin{tabular}{c c c c c}        
    \hline         
    Wavelength(\AA) & $P (\%)$ & PA ($^{\circ}$) \\  
    \hline
    4440 - 5105 & $0.11 \pm 0.01$ & $13.2^{\circ} \pm 2.7^{\circ}$ \\\relax
    5105 - 5770 & $0.09 \pm 0.009$ & $9.1^{\circ} \pm 2.7^{\circ}$ \\\relax
    5770 - 6505 & $0.09 \pm 0.008$ & $8.4^{\circ} \pm 2.6^{\circ}$ \\\relax
    6505 - 6620 (H$\alpha$ + [N II]) & $1.29 \pm 0.13$ & $0.5^{\circ} \pm 3.0^{\circ}$ \\\relax
    6620 - 6700 & $0.09 \pm 0.03$ & $7.7^{\circ} \pm 8.5^{\circ}$ \\\relax
    6700 - 6750 ([S II]) & $1.18 \pm 0.53$ & $-0.6^{\circ} \pm 12.9^{\circ}$ \\\relax
    6750 - 7440 & $0.08 \pm 0.009$ & $6.6^{\circ} \pm 3.1^{\circ}$ \\\relax
    7440 - 8105 & $0.06 \pm 0.01$ & $19.1^{\circ} \pm 4.8^{\circ}$ & \\\relax
    8105 - 9260 & $0.07 \pm 0.01$ & $18.1^{\circ} \pm 3.9^{\circ}$ \\
    \hline                                          
    \end{tabular}
\end{table}

\subsection{Modeling the polarized emission lines with narrow lines only}
\begin{figure}
    \centering
    \includegraphics[width=0.9\linewidth]{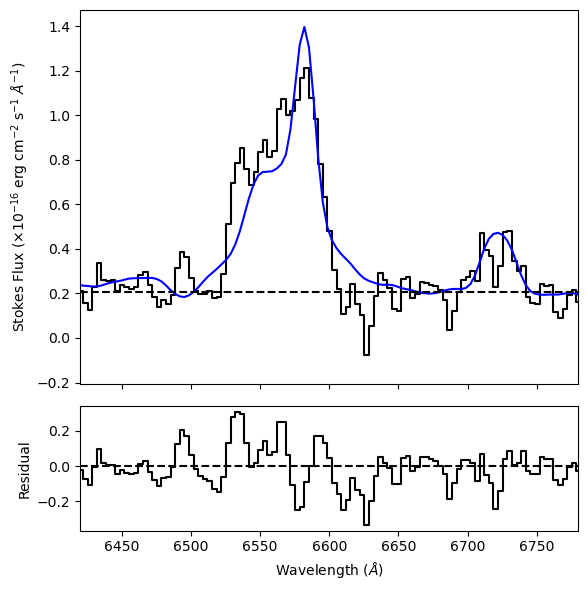}
    \caption{The upper panel shows modeling the polarized line profile (in black) with the profile from the total flux spectrum (in blue). The lower panel shows the residuals from the fitting.}
    \label{fig:fitting}
\end{figure}

We first examine if the scattered narrow line emissions solely can account for the features in polarized flux spectrum. To this end, we model the Stokes flux profiles of the H$\alpha$ and [S~II] complexes using the exact profiles extracted from the total flux spectrum, varying only the normalization. As shown in Fig.~\ref{fig:fitting}, modeling the polarized line profiles with only narrow lines provides an acceptable fit ($\chi_{\nu}^{2} = 1.6)$ as far as feature width is concerned. Importantly, no broad component is required in the Stokes flux spectrum to account for underlying emission from the BLR. We acknowledge that the profile in the Stokes flux and total flux may differ in detail since the scattered narrow lines see different parts of the mirror when the size of the mirror is comparable with the NLR. However, only non-unique, ad hoc explanations can be given for this peculiar detail.

Traditional methods that place an upper limit on the broad H$\alpha$ flux from the BLR require assuming the profile of the undetected line, which can lead to erroneous constraints (e.g., NGC~3147, \citealt{Bianchi2019, Bianchi2022}; see also the discussion in \citealt{Antonucci2012}). Therefore, in the next section, we adopt a different approach, aiming not to rigorously constrain the H$\alpha$ contribution from the BLR, but to robustly rule out the presence profiles typical of Seyfert~I AGN.

\subsection{Precluding Seyfert I type line profile in the polarized flux spectrum} 
Is it possible to robustly rule out the possibility that the polarized flux contains a Seyfert I-type line profile? We adopt the method outlined in \citet{Marin2025} to evaluate what fraction of the genuine type-I profiles would be conspicuous if present in our polarized flux spectrum.

Accurate determination of the polarized continuum flux is essential since the polarization of the continuum is heavily diluted by starlight in total flux. Using HST imaging polarimetry, where significant polarization has been detected \citep{Capetti2007}, we obtain a polarized flux of $\simeq 2.6 \times 10^{-17} $\,erg\,s$^{-1}$\,cm$^{-2}$\,\AA$^{-1}$ within 1.2$^{\prime\prime}$ radius aperture, comparable to the average seeing of our observations. Given that polarized narrow emission lines contribute approximately 20\% to this flux with the same polarization angle, the polarized continuum flux is $\simeq 2.0 \times 10^{-17}$ erg\,s$^{-1}$\,cm$^{-2}$\,\AA$^{-1}$, in excellent agreement with our observations. 

We scale the real type-I profiles onto the polarized continuum based on their equivalent widths and resample the coadded spectra according to the uncertainties in our polarized flux. The type-I AGN sample is drawn from \citet{Schmidt1983} (23 sources), \citet{Stripe1990} (5 sources), and Jiang et al. (in prep.; 9 sources). The properties of these sources are summarized in Table \ref{tab:BQS_sample}. This sample comprises highly luminous objects ($M_{B}/M_{V} < -23.0$\,mag or log($L_{5100}$/erg\,s$^{-1}$\,cm$^{-2}) > 45.0$), whose spectra can be compared directly to our polarized flux spectrum due to the absence of host stellar contamination \citep{Shen2011}. We assume that this sample is representative of classified type-I AGNs. We calculate the SNR of the broad wing region in the coadded spectra, defined at the wavelength corresponding to the half-maximum. A broad feature is considered detected if the SNR at this wavelength exceeds 3. We find that fewer than 6\% (2 out of 37, see Fig.~\ref{fig:scaled_flux}) of the type-I profiles would remain undetected.\\

Therefore, we conclude that the presence of scattered narrow lines is more plausible in 3C~270, and neither a BLR nor a BBB is required under these circumstances. 

\section{Discussion and Summary}
\label{sec:discussion}
\citet{Chiaberge1999} studied a sample of FR~I radio galaxies using the HST and identified an unresolved optical Compact Central Core (CCC) with a high detection rate, including 3C~270. The presence of a CCC in 3C~270 suggests that there is only mild extinction $A_V$ ($\lesssim$ 3) along the line of sight, as substantial extinction would completely obscure the source \citep{Chiaberge2003}.  Typically, low extinction corresponds to a low column density $N_\mathrm{H}$ and further examination on the X-ray emission shows that for most of the sources in the sample, $N_\mathrm{H}$ is indeed very low \citep[$\lesssim 0.5 \times 10^{21}$\,cm$^{-2}$,][]{Balmaverde2006}. However, X-ray studies of 3C~270 require modeling the spectrum with a significantly higher column density $N_\mathrm{H}$ of $ \gtrsim 7 \times 10^{22}$\,cm$^{-2}$ \citep{Zezas2005, Worrall2010}. 

The observed discrepancy between high column density and low extinction in 3C~270 could be explained if most of the X-ray obscuring material is not along the line of sight to the optical emission. In our favored scenario, we are observing the optical emission through a mirror and do not expect X-ray column density to apply.

If the high X-ray column density does mean the optical core is hidden in this object, our data indicate that we can see the hidden nuclear region with a polar mirror.  Since we don't observe broad lines in polarized flux but we do observe continuum and narrow lines, it seems likely that there is no hidden BBB and BLR. Thus, our periscopic view reveals the underlying continuum, including any contributions from Radiatively Inefficient Accretion Flow (RIAF) and beamed synchrotron, and the innermost NLR. With the available data, we can not distinguish which of the two components dominates the scattered continuum. In the simplest FR~I\,–\,BL~Lac unification scheme \citep{Chiaberge1999, Capetti2000, Trussoni2003}, a scattered continuum dominated by beamed synchrotron emission would be favored. \citet{deMenezes2020} attempted a broadband SED analysis of 3C~270 and favored a jet-dominated synchrotron self\,-\,Compton model over the RIAF model. However, we caution that the optical emission being scattered should be fully considered when performing such analyses, and that multi-wavelength (polarimetric) SED analysis are needed to better distinguish the two components \citep[e.g.,][]{Marin2018}. In either cases, the presence of a mid-infrared point source \citep{Asmus2011, Asmus2014} and warm dust emission \citep{Leipski2009} in 3C~270 arising from the BBB heating the compact torus can be ruled out. This is our favored hypothesis, making 3C~270 similar to radio-loud AGN Centaurus~A \citep{Marin2025} and radio-quiet AGN NGC~4258 \citep{NGC4258}. The caveat is that the scattered continuum is only weakly detected, so a commensurate BBB and BLR can't be totally ruled out.

\vspace{5mm}
\begin{acknowledgments}
We thank the anonymous referee for their thoughtful feedback and constructive comments. The authors acknowledge M. J. Martínez González for performing the data reduction and preliminary analysis and C. Ramos Almeida for useful discussions.
\end{acknowledgments}

\vspace{5mm}
\facilities{VLT (FORS2)}

\software{Astropy \citep{Astropy2013, Astropy2018}, }

\appendix
\renewcommand{\thetable}{A\arabic{table}} 
\setcounter{table}{0} 
\begin{table}
    \label{tab:BQS_sample}      
    \centering        
    \begin{tabular}{c c c c c}        
    \hline             
    Quasar ID & Redshift & Selection Criteria & Reference \\  
    \hline
    PG 0157+001 & 0.164 & $M_{B}$ = -24.83 & 1, 3\\\relax
    PG 0923+201 & 0.190 & $M_{B}$ = -24.32 & 1, 3\\\relax
    PG 0947+396 & 0.206 & $M_{B}$ = -24.15 & 1, 3\\\relax
    PG 1001+054 & 0.161 & $M_{B}$ = -23.86 & 1, 3\\\relax
    PG 1004+130 & 0.240 & $M_{B}$ = -24.96 & 1, 3\\\relax
    PG 1012+008 & 0.185 & $M_{B}$ = -24.41 & 1, 3\\\relax
    PG 1048+342 & 0.167 & $M_{B}$ = -24.26 & 1, 3\\\relax
    PG 1114+445 & 0.144 & $M_{B}$ = -23.69 & 1, 3\\\relax
    PG 1115+407 & 0.154 & $M_{B}$ = -23.87 & 1, 3\\\relax
    PG 1116+215 & 0.177 & $M_{B}$ = -25.03 & 1, 3\\\relax
    PG 1202+281 & 0.165 & $M_{B}$ = -25.03 & 1, 3\\\relax
    PG 1229+204 & 0.064 & $M_{B}$ = -23.30 & 1, 3\\\relax
    PG 1307+085 & 0.155 & $M_{B}$ = -24.63 & 1, 3\\\relax
    PG 1309+355 & 0.184 & $M_{B}$ = -24.84 & 1, 3\\\relax
    PG 1352+183 & 0.158 & $M_{B}$ = -24.24 & 1, 3\\\relax
    PG 1404+226 & 0.098 & $M_{B}$ = -23.07 & 1, 3\\\relax
    PG 1415+451 & 0.114 & $M_{B}$ = -23.48 & 1, 3\\\relax
    PG 1427+480 & 0.221 & $M_{B}$ = -24.37 & 1, 3\\\relax
    PG 1519+226 & 0.137 & $M_{B}$ = -23.54 & 1, 3\\\relax
    PG 1545+210 & 0.266 & $M_{B}$ = -25.08 & 1, 3\\\relax
    PG 1552+085 & 0.119 & $M_{B}$ = -23.30 & 1, 3\\\relax
    PG 1612+261 & 0.131 & $M_{B}$ = -23.53 & 1, 3\\\relax
    PG 2233+134 & 0.326 & $M_{B}$ = -25.54 & 1, 3\\\relax
    PKS 1020-103 & 0.197 & $M_{V}$ = -23.53 & 2\\\relax
    Fairall 9 & 0.046 & $M_{V}$ = -23.25 & 2\\\relax
    PG 0837-12 & 0.198 & $M_{V}$ = -24.13 & 2\\\relax
    4C 31.63 & 0.295 & $M_{V}$ = -25.31 & 2\\\relax
    LEDA 61965 & 0.123 & $M_{V}$ = -23.30 & 2\\\relax
    SDSS J014017.06-005003.0 & 0.334 & log($L_{5100}$/erg s$^{-1}$) = 45.19& 3\\\relax
    SDSS J023922.87-000119.6 & 0.262 & log($L_{5100}$/erg s$^{-1}$) = 45.12& 3\\\relax
    SDSS J082758.52+233302.9 & 0.302 & log($L_{5100}$/erg s$^{-1}$) = 45.12& 3\\\relax
    SDSS J082958.79+513044.5 & 0.342 & log($L_{5100}$/erg s$^{-1}$) = 45.17& 3\\\relax
    SDSS J083535.80+245940.1 & 0.330 & log($L_{5100}$/erg s$^{-1}$) = 45.23& 3\\\relax
    SDSS J100402.60+285535.3 & 0.329 & log($L_{5100}$/erg s$^{-1}$) = 45.30& 3\\\relax
    SDSS J143204.60+394439.0 & 0.348 & log($L_{5100}$/erg s$^{-1}$) = 45.09& 3\\\relax
    SDSS J151535.25+480530.5 & 0.312 & log($L_{5100}$/erg s$^{-1}$) = 45.05& 3\\\relax
    SDSS J161849.24+442517.2 & 0.335 & log($L_{5100}$/erg s$^{-1}$) = 45.03& 3\\
    \hline
    \end{tabular}
    \caption{The sample consists of 37 luminous quasars. (1) The absolute B-band magnitudes are adopted from \citet{Schmidt1983}; (2) the absolute V-band magnitudes are taken from \citet{Stripe1990} and corrected for the adopted cosmological parameters; (3) the spectra are retrieved from the SDSS SkyServer (\url{https://skyserver.sdss.org/dr18}).}
\end{table} 
\clearpage

\renewcommand{\thefigure}{A\arabic{figure}}
\setcounter{figure}{0} 
\begin{figure*}
    \centering
    \includegraphics[width=0.9\linewidth]{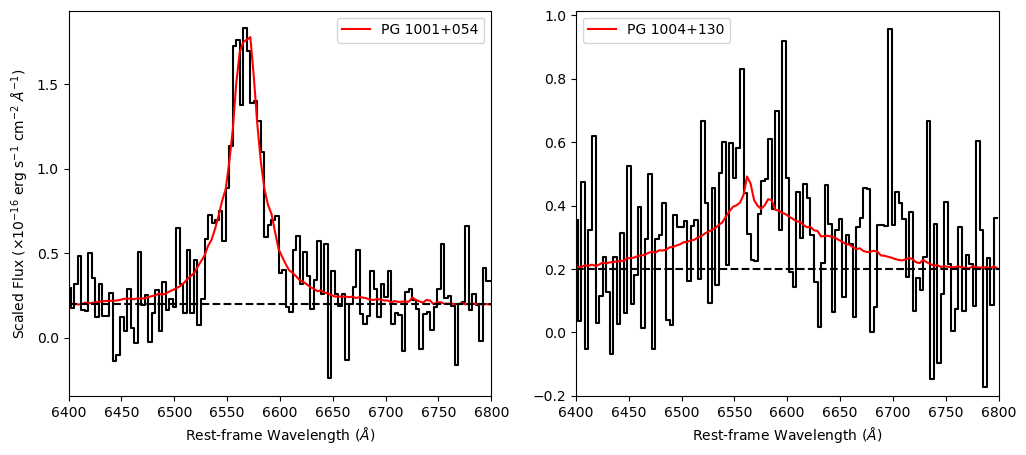}
    \caption{The left panel illustrates the case where the scaled type-I profile from PG 1001+054 is significantly detected, while the right panel demonstrates the scenario for PG 1004+130, where the broad line profile is not confidently detected. In each panel, the red line represents the scaled original line profile of PG 1001+054 and PG 1004+130, adjusted based on its equivalent width. The black line shows the scaled line profile further resampled according to the uncertainty in our polarized flux spectrum.}
    \label{fig:scaled_flux}
\end{figure*}

\bibliography{citation}{}
\bibliographystyle{aasjournal}



\end{document}